\documentstyle[prd,aps]{revtex}

\begin{document}


\title{
\null
\vskip-6pt \hfill {\rm\small MCTP-02-24} \\
\vskip-6pt \hfill {\rm\small CU-TP-1059} \\
\vskip-9pt~\\
The Ultimate Fate of Life in an Accelerating Universe}

\vspace{.5in}

\author{Katherine Freese$^{1,2}$\thanks{Electronic address: {\tt ktfreese@umich.edu}} and William H. Kinney$^2$\thanks{Electronic address: {\tt kinney@physics.columbia.edu}}}

\address{
\vspace{.7cm}
$^1$Michigan Center for Theoretical Physics,
University of Michigan,
Ann Arbor, MI 48109, USA\\
$^2$Institute for Strings, Cosmology and Astroparticle Physics, 
Columbia University,
550 W. 120th St., New York, NY 10027\\
}

\maketitle

\begin{abstract}

The ultimate fate of life in a universe with accelerated expansion
is considered.  Previous work \cite{barrowtipler,starkkrauss} showed
that life cannot go on indefinitely in a universe dominated
by a cosmological constant.  In this paper we consider instead
other models of acceleration (including quintessence and Cardassian
expansion).  We find that it is possible in these cosmologies for life to 
persist indefinitely.  As an example we study potentials of the
form $V \propto \phi^n$ and find the requirement $n < -2$.

\end{abstract}
\pacs{98.80.-k,89.70.+c}


\section{Introduction}

In 1979, Dyson first discussed the question of the ultimate
fate of life in an expanding universe\cite{dyson}.  He proposed a framework
within which to discuss whether or not some form of life,
material or otherwise, can go on.  At the time
of his work the universe was assumed to be decelerating.
However, in the light of recent evidence that the universe
is accelerating, the conclusions of Dyson's original work
deserve reinvestigation.  Observations of Type IA Supernovae \cite{SN1,SN2}
as well as concordance with other observations (including the microwave
background and galaxy power spectra) indicate that
the universe is accelerating.   Hence the question of the future
of life in our universe deserves another look in the context
of this acceleration.  Barrow and Tipler\cite{barrowtipler} and 
 Krauss and Starkman\cite{starkkrauss} followed
the basic approach outlined by Dyson to consider 
life in a universe dominated by a cosmological constant.
They concluded that life is inevitably doomed to oblivion in such
a universe.  Any lifeform would eventually fry to death
in the bath of thermal Hawking radiation produced by the
de Sitter vacuum\cite{starkkrauss}.  
Beings of any kind generate heat by the process
of living and eventually are unable to dissipate their heat in the
background of this thermal bath.

In this paper we consider the consequences of other explanations
for the acceleration of our universe.  Other than a cosmological
constant, alternatives include
a decaying vacuum energy \cite{fafm}\cite{frieman},
quintessence\cite{PR,Wett,WCL,FJ,CDS,CLW,ZWS,BM,BCN,Arm}, and
Cardassian expansion \cite{cardass} as possible explanations
for such an acceleration. Quintessence has a time dependent vacuum energy
given by a rolling scalar field. Cardassian expansion is a model
with matter and radiation alone (no vacuum at all) in which acceleration is
driven by a modified Friedmann Robertson Walker (FRW) equation.
The crucial difference between these
cases and that of a cosmological constant 
is that the temperature of the cosmological Hawking
radiation decreases in time, in many cases quickly enough to allow
life to continue indefinitely despite the presence of the thermal
bath. We consider two cases: (1) a constant equation of state
$p = w \rho$ with $-1 \leq w < -1/3$ (which includes the case
of Cardassian expansion for constant exponent $n$ defined below), 
and (2) a time-varying
equation of state generated by a ``quintessence'' potential 
of the form $V\left(\phi\right) \propto \phi^{n}$ with $n < 0$. In the 
case of constant equation of state, we find that any equation 
of state {\em except} a cosmological constant ($w = -1$) allows
for the indefinite continuation of life. In the quintessence case,
we find that any potential $V \propto \phi^{n}$ with $n < -2$ is
consistent with the indefinite continuation of life. 

In the conclusions, we discuss speculative scenarios in which life
might avoid inevitable extinction even in the case of a universe
dominated by a cosmological constant, including quantum computation,
oscillating universes, wormholes, and laboratory-created universes. We
also comment on the argument of Krauss and Starkman that the presence
of a quantum-mechanical ground state for the system renders Dyson's
argument invalid in general. We argue that the inclusion of a
cosmological Hawking temperature in Dyson's classical argument correctly
captures the quantum nature of the system and that therefore Dyson's
conclusions are in fact valid.

\section{The premise set out by Dyson:}

Dyson introduced the ``biological scaling'' hypothesis to estimate the
rate at which an organism in an environment of temperature $T$ can 
perform computation. We refer the reader to Ref. \cite{dyson} for a
definition and detailed discussion of the scaling hypothesis, and 
instead concentrate here on its relevant consequences. The important 
consequence of Dyson's 
scaling hypothesis is  the notion of ``subjective time'', i.e.,
the appropriate measure of time as experienced by a living creature
is the quantity
 \begin{equation}
\label{eq:comp}
u(t) = f\int_{t_0}^t T(t') dt' \label{eqsubjtime1}
\end{equation}
where $T(t)$ is the temperature of the creature and
$f=(300$deg sec)$^{-1}$ is introduced so as to make $u$ 
dimensionless\footnote{The value of $f$ suggested by Dyson 
is motivated by the fact
that human life takes place at 300K with each moment of consciousness
lasting about a second; the precise value of $f$ is immaterial to the
arguments.}. One can think of the quantity $u$
as the number of possible computations in a time $t$. We can define
one ``computation'' as some change of state in a quantum system. Then
a single computation, from the energy/time uncertainty relation, takes 
place over a characteristic timescale
\begin{equation}
\Delta t \geq {\hbar \over \Delta E} \sim {\hbar \over k T},
\end{equation}
where $k$ is Boltzmann's constant. Then the total number of computations
$\Delta u$ over time $\Delta t$ is then given by
\begin{equation}
{\Delta u \over \Delta t} \propto {k T \over \hbar},
\end{equation}
which leads directly to the definition of subjective time in Eq.
(\ref{eqsubjtime1}). (Later we comment on the possibility of 
using quantum computation to alter this notion of subjective time.) The 
continuation of life requires the possibility of an infinite number of 
computations in a system with only a finite amount of energy.  

Dyson points out a second consequence of the scaling law: any creature
is characterized by a quantity $Q$ which measures its rate
of entropy production per unit of subjective time, $dS = Q du$, in some 
sense the ``complexity'' of the creature.
Dyson estimates that, for a human
dissipating about 200W of power at 300K, $Q \sim 10^{23}$ bits.
Krauss and Starkman estimate that the uncertainty in this number
suggests that  a civilization of conscious beings requires
${\rm log}Q >50-100$.
Any creature in the process of living and computing will
dissipate energy.
 A lifeform with given $Q$ and given temperature $T$ will dissipate
energy at a rate
\begin{equation}
\label{eq:meta}
m \equiv {dE \over dt} = k T {dS \over dt} = k f Q T^2,
\end{equation}
where $m$ is the metabolic rate measured in ergs per second.
The total energy consumed by the creature is then
\begin{equation}
E = k f Q \int_{t_0}^t T^2(t') dt'.\label{eqenergy1}
\end{equation}
Since the rate of computation scales as $T$ while the rate of energy
consumption scales at $T^2$, it at first appears possible that an organism
can perform an infinite number of computations using a finite amount
of energy, as long as the operating temperature of the organism
continuously decreases in time, $T(t) \propto t^{-\alpha}$, with 
$1/2 < \alpha \leq 1$. We will refer to this as {\em Dyson's condition}: 
life can be considered ``infinite'' if the number of computations, or
``subjective time'', (\ref{eqsubjtime1}) can be infinite while the total
energy consumed (\ref{eqenergy1}) is finite. 

This naive analysis assumes that the organism is completely free to 
choose its temperature $T(t)$ so as to satisfy Dyson's condition.
Several constraints restrict this temperature.
The creature must be able to get rid of the heat $E$ generated by
the computations it performs (\ref{eq:meta}). However, Dyson estimates 
an upper limit to the rate at which waste heat can be radiated as
\begin{equation}
I(t) < 2.84 {N_e e^2 \over m_e \hbar^2 c^3} \left(k T\right)^3.
\end{equation}
The creature will fry to death unless it can dissipate the heat $E$
that it creates; dissipation by radiation implies a lower limit
on the operating temperature for the organism:
\begin{equation}
T(t) > T_{\rm min} = (Q/N_e) 10^{-12} K.\label{eqTmin1}
\end{equation}
Since the ratio $(Q/N_e$) between the complexity of the society and
the number of electrons at its disposal cannot be made arbitrarily
small, there must be a finite minimum temperature for which computation
is possible. Therefore Dyson's condition cannot be satisfied, and 
the creature or society cannot survive indefinitely.

However, Dyson proposes a strategy to avoid this sad conclusion:
hibernation.  Life may find a way to metabolize intermittently,
yet continue to radiate waste heat into space during its
periods of hibernation.  The society can remain active for  a fraction
$g(t)$ of its time while hibernating for the remaining $1-g(t)$
fraction of the time.
During these periods of hibernation,
metabolism can be effectively stopped while radiation of waste heat
continues.
Then the total subjective time is modified to 
\begin{equation}
u(t) = f \int_0^t g(t') T(t') dt', \label{eqsubjtime2}
\end{equation}
while the average rate of heat production by the organism becomes
\begin{equation}
m= k f Q g T^2 .
\end{equation}
Therefore the temperature of the organism can drop below $T_{\rm min}$ 
in Eq. (\ref{eqTmin1}) and the heat generated by the computation can 
still be dissipated: the condition (\ref{eqTmin1}) becomes
\begin{equation}
\label{eq:disshib}
T(t) > T_{\rm min} \equiv {Q \over N_e} g(t) 10^{-12} K.\label{eqTmin2}
\end{equation}
As long as the operating temperature of the organism is above this limit,
it can dispose of waste heat. The total energy consumed is
\begin{equation}
E = k f Q \int_{t_0}^{t}{g(t') T^2(t') dt'}. \label{eqenergy2}
\end{equation}
The organism is free to choose $g(t)$ and $T(t)$ to satisfy Dyson's 
condition. We will assume (consistent with other authors) that
$g(t) \propto T(t) \propto t^{-p}$, the minimum amount of hibernation
consistent with the energy dissipation condition (\ref{eqenergy2}). 
Then the subjective time is given by
\begin{equation}
u(t) \propto \int_{t_0}^t{(t')^{-2 p} dt'},
\end{equation}
and the total energy consumed scales as
\begin{equation}
E(t) \propto \int_{t_0}^t{(t')^{-3 p} dt'}.
\end{equation}
Dyson's condition
\begin{eqnarray}
u(t \rightarrow \infty) &&\rightarrow \infty,\cr
E(t \rightarrow \infty) &&\rightarrow {\rm const.}
\end{eqnarray}
is then satisfied for 
\begin{equation}
\label{eq:p}
1/3 < p \leq 1/2.
\end{equation}
An additional constraint is generated by the fact that
the creatures, even if hibernating, cannot cool off any
faster than the background universe.  Hence, if the 
universe temperature scales as
\begin{equation}
T_u(t) \propto t^{-q},
\end{equation}
then this second constraint requires that
\begin{equation}
\label{eq:back}
p<q .
\end{equation}

It is clear that Eqs.(\ref{eq:p}) and (\ref{eq:back}) can
both be simultaneously satisfied
in a (decelerating) Cold Dark Matter-dominated cosmology.  There the 
temperature of the background universe is
given by the  Cosmic Microwave Background
temperature, which scales in a matter-dominated cosmology at 
$T_{\rm CMB} \propto t^{-2/3}$, i.e., $q = 2/3$.
Hence the background temperature indeed drops 
more quickly than the temperature required 
for the organism to satisfy Dyson's bound.  

\section{Cosmological Constant Dominated Universe}

Krauss and Starkman considered modifications to these
questions in the context of a universe dominated by a
cosmological constant $\Lambda$. 
In de Sitter space, Hawking radiation
creates a thermal bath of particles at the de Sitter temperature,
\begin{equation}
T_{deS} = \sqrt{\Lambda \over 12 \pi^2} = {\rm constant}.
\end{equation}
Hence the universe itself has a fundamental minimum temperature,
with $q=0$.  Then the 
constraint in Eq.(\ref{eq:disshib}) is replaced by
\begin{equation}
\label{eq:both}
T(t) > T_{min} \equiv {\rm max}
\left[T_{deS}, {Q \over N} g(t) 10^{-12}K\right].
\end{equation}
Therefore no hibernation strategy will be sufficient to satisfy 
Dyson's condition. The first term in Eq.(\ref{eq:both})
eventually dominates, and then
Eq.({\ref{eq:back}}) cannot be satisfied with $q=0$.
Eventually thermal equilibrium will be reached
with everything at the Hawking temperature of the thermal bath, and
further computation will be an impossibility. Life in a universe with
a cosmological constant is doomed to extinction.

\section{Dark energy}

The ``dark energy'' driving the acceleration 
of the universe, however, need not be a cosmological constant. 
In a more general scenario, the energy density driving 
the acceleration can be variable in time, or, equivalently, 
have an equation of state $p > -\rho$. Acceleration takes place for any equation of state $p < -1/3 \rho$. In this section, we examine the case of a more general equation of state and show that Dyson's condition for infinite computation can be met for a wide range of accelerating cosmologies. 

\subsection{Constant Equation of State}

We first consider the simple case of equation of state $p = w \rho$, with $w$ constant in time. Any accelerating cosmology evolves toward flatness at late time, so we can assume a flat cosmology. From the Friedmann equation
\begin{equation}
H^2 \equiv \left({\dot a \over a}\right)^2 = 
{8 \pi \over 3 m_{\rm Pl}^2} \rho
\end{equation}
and the Raychaudhuri equation
\begin{equation}
\left({\ddot a \over a}\right) = - {4 \pi \over 3 m_{\rm Pl}^2} \left(\rho + 3 p\right),
\end{equation}
we have, for $w = {\rm const.}$,
\begin{equation}
{d H \over d t} = - {3 \over 2} \left(1 + w\right) H^2.
\end{equation}
For the case of a cosmological constant, $w = -1$ 
and $H = {\rm const.}$, so that $q=0$ and one can never satisfy
Eq.(\ref{eq:back}). 
Therefore Dyson's condition is violated. 
However, for $w > -1$, we have $H \propto t^{-1}$. 
The Hawking temperature of the space (the generalization of the
de Sitter temperature in de Sitter Space)
therefore also decreases as $T_{\rm H} \propto H \propto t^{-1}$, 
i.e., $q=1$.  Then, in Eq.(\ref{eq:back}), the first
term ($\propto t^{-1})$ drops more rapidly than the second
($\propto t^{-p}$),
and one is back to Dyson's original condition in Eq.(\ref{eq:disshib}).
As long as Eq.(\ref{eq:p}) is satisfied,
the Dyson condition that an infinite amount of 
computation be possible with finite total energy expended 
can be met in any cosmology with $w = {\rm const.}$ {\em except} 
the special case of a cosmological constant, $w = -1$. 

\subsection{Time Varying Equation of State}
In general, however, the equation of state of the dark energy 
need not remain constant. We cannot comment in general upon
all time-varying equations of state.  We here concentrate
upon the particular case of ``quintessence'' models,
in which the dark energy consists of a slowly rolling scalar field.
The time-dependence of the equation of state depends 
on the form of the potential for the quintessence field $\phi$. The equation of motion for a scalar field in a cosmological background is
\begin{equation}
\ddot \phi + 3 H \dot \phi + V'\left(\phi\right) = 0.\label{eqscalareom}
\end{equation}
The case of an exponential potential, $V\left(\phi\right) \propto \exp\left(\phi / M\right)$ is just that of a constant equation of state considered above, since
\begin{equation}
a\left(t\right) \propto t^{1 / \epsilon},
\end{equation}
with $\epsilon = {\rm const.} < 1$ corresponding to accelerated expansion. The equation of state is
\begin{equation}
w = {2 \over 3} \epsilon - 1 = {\rm const.}
\end{equation}
This example can be generalized 
to an arbitrary potential as follows: For a slowly rolling scalar field $\phi$, the equation of motion (\ref{eqscalareom}) is approximately
\begin{equation}
3 H \dot\phi \simeq - V'\left(\phi\right),
\end{equation}
and the Friedmann equation is
\begin{equation}
H^2 = {8 \pi \over 3 m_{\rm Pl}^2} \left[{1 \over 2} \dot \phi^2 + V\left(\phi\right)\right] \simeq  {8 \pi \over 3 m_{\rm Pl}^2} V\left(\phi\right).
\end{equation}
Then
\begin{eqnarray}
{d H \over d t} &=& {1 \over 2} H \left({V'\left(\phi\right) \over V\left(\phi\right)}\right) \dot \phi\cr
&=& - \epsilon H^2,
\end{eqnarray}
where the slow-roll parameter $\epsilon$ is 
\begin{equation}
\epsilon = {m_{\rm Pl}^2 \over 16 \pi} \left({V'\left(\phi\right) \over V\left(\phi\right)}\right)^2.
\end{equation}
The exponential potential then has $\epsilon = {\rm const.}$ as above, but $\epsilon$ varies with time for arbitrary potential. The equation of state
\begin{equation}
w \simeq {2 \over 3} \epsilon - 1,
\end{equation}
therefore varies in time as well. We consider the class of potentials
\begin{equation}
V\left(\phi\right) \propto \phi^n.
\end{equation}
Quintessence models generally take $n < 0$, so that the universe is accelerating in the late-time limit. The slow-roll parameter $\epsilon$ is then
\begin{equation}
\epsilon = {n^2 m_{\rm Pl}^2 \over 16 \pi} \left({1 \over \phi^2}\right) \propto H^{-4 / n}.
\end{equation}
The equation of motion for the Hubble parameter is
\begin{equation}
{d H \over d t} \propto - H^{2 (n - 2) / n},
\end{equation}
with solution
\begin{equation}
H\left(t\right) \propto \left[\left({n - 4 \over n}\right) t + {\rm const.}\right]^{n / \left(4 - n\right)}.
\end{equation}
We are interested in the solution at late times, 
so that for the quintessence case of $n < 0$, the Hubble 
parameter and the Hawking temperature evolve as
\begin{equation}
T_{\rm H} \propto H \propto t^{-\left\vert{n / \left(4 - n\right)}\right\vert}
\end{equation}
so that $q = \left\vert n /  (4 - n)\right\vert$.
The Dyson condition requires $p<q$ 
as $t \rightarrow \infty$, where we are allowed to choose $p$ anywhere in the range $1/3 < p \leq 1/2$. Taking the {\em slowest} rate of falloff for $T_{\rm min}$, we have the condition
\begin{equation}
\left\vert{n \over 4 - n}\right\vert > p > 1/3,
\end{equation}
so that Dyson's condition can be satisfied for quintessence models with $n < -2$. 

\section{Modified FRW Equations}

An alternative way to drive acceleration of the universe is modification
of the Friedmann Robertson Walker equations.
In Cardassian expansion \cite{cardass}, the FRW equations become
\begin{equation}
H^2 = {8 \pi \over 3 m_{pl}^2} \rho + B \rho^n
\end{equation}
with $n<2/3$.  In this model there is no vacuum term at all,
and the energy density $\rho$ is simply given by ordinary matter and 
radiation.  The second term becomes more important as time goes on
and for redshifts $z<1/2$ drives acceleration of the universe.
The universe is thus flat, matter dominated, and accelerating in this model.
An alternate way to modify the FRW equations has been studied by
\cite{ddg}.
For constant coefficient $n$ in the Cardassian model, 
the background evolution of the 
universe behaves dynamically the same as a constant equation of state
$w=n-1$ so that the conclusion in the previous section implies
that life can persist in a universe with $n>0$.

\section{Conclusion and Discussion}

We have found that life can go on indefinitely in an accelerating
universe, depending on what energy density drives the acceleration.
Previous authors\cite{barrowtipler,starkkrauss} showed that the 
(time independent) de Sitter
radiation in a cosmological constant driven expansion
destroys all life eventually.  But in other cases we
considered, including quintessence and Cardassian expansion,
we found that the de Sitter radiation cools off just rapidly
enough to allow life to survive. In particular, for any 
constant effective equation of state $w = p/\rho > -1$,
and any constant Cardassian exponent $n> 0$,
life can persist.  In addition, we considered time-varying
equations of state for the case of a quintessence with
potential $V \propto \phi^n$, and found successful futures
for life if $n < -2$.  

In this work we followed the basic premise set up by Dyson.
Currently we understand that there is a disagreement between
Dyson on the one hand and Krauss and Starkman (KS)
on the other hand as to whether or not there
is a flaw in the premise.  KS argue that any system in
which computation is an irreversible process must eventually reach
a quantum-mechanical ground state, beyond which further metabolism
will not be possible. If such a system is finite, it must necessarily
reach the ground state in finite time. We argue that this line of
reasoning is valid only in the limit of a {\em static} (i.e., de Sitter) 
spacetime. Consider the phase space available for quantum modes in
an accelerating spacetime. The horizon size $d_{\rm H} \sim H^{-1}$
provides an infrared cutoff, since modes with momentum $p < H$
have wavelength longer than the horizon size, at which point they
become classical perturbations. Therefore the horizon size defines
an effective ground state for quantum modes in the spacetime, $E > H$.
However, this is exactly the physics which leads to the Hawking 
temperature $T_{\rm deS} \sim H$! In the case of exact de Sitter
space, the ground state energy is constant in time, and therefore
the argument of KS that any finite system must relax to its ground
state in finite subjective time is valid. However, in backgrounds where 
the horizon size is increasing in time, the ``ground state'' energy
defined by the infrared cutoff is decreasing in time and the
system continuously has new, lower-energy states made available to
it. Classically, this behavior is manifest in the time-dependence
of the Hawking temperature. Thus the system can continue to radiate
waste heat and reaches a ground state only after infinite subjective
time, exactly as suggested by the classical calculation. This argument
is obviously speculative, and it would be desirable to frame it in
a more quantitative way. In particular, it is not clear that a system
with these properties can truly be considered ``finite''.

One might wonder if quantum computing would allow us to modify the Dyson 
condition in a useful way.  Then the number of computations (the 
``subjective time'') given in Eq.(\ref{eqsubjtime1}) will be much larger 
for a given rate of energy dissipation.  A lifeform
may clearly continue to live or compute for a much longer time period
with the same energy consumption. However it is 
straightforward to show that including quantum computation as a possibility  
does not affect our conclusions about the ultimate fate of life. 
The above discussion was based on the 
thermodynamics of a conventional computer, which uses energy at a 
rate of $m = f Q k T^2$ 
to ``flip'' $Q$ bits at temperature $T$. We can make an optimistic estimate 
of the increase in efficiency afforded by quantum computing by supposing that
any operation performed on $Q$ bits by a conventional computer can
be performed on a superposition of $2^Q$ entangled quantum states by a
quantum computer, with identical energy consumption. Thus a classical
system with complexity $Q$ can be built as a quantum system with complexity
$\log_2(Q)$, which dissipates energy at a rate
\begin{equation}
m_{\rm quant} = f \log_2(Q) k T^2.
\end{equation}
However, this improvement in efficiency alters the energy integral 
(\ref{eqenergy2}) by a multiplicative factor, and has no effect on 
whether the total energy consumed is finite or infinite. Therefore
our arguments apply equally well to quantum as well as classical 
computers. However, we note that an organism of a given complexity
$Q$ can live exponentially longer in subjective time by adopting
quantum computing as a strategy.

We note that a finite system, while it may be capable of an infinite
amount of computation, is only capable of storing a {\em finite} number
of memories. As long as the expansion of the universe is accelerating, any 
system which is initially finite must remain so, since any additional 
material with which to build new ``memory'' has redshifted beyond the 
horizon and is therefore unavailable. We thus reach the apparently 
inescapable  conclusion that, while life itself may be immortal, any 
individual is doomed to mortality.

There are certainly limitations to Dyson's premise.  One alternative
cosmology which would violate
Dyson's premise would be if the universe oscillates \cite{tolman,freesekinney,freesestarkman} or is cyclic \cite{steintur}.  Then the current
accelerating phase might be followed by a subsequent recontraction
and then again an expansion, and life could begin all over again.
Of course the new burst of life might not have any memory
of our current cycle, so that this does not provide an altogether
satisfactory solution to the problem of enabling life to continue
indefinitely\cite{dyson}.

However, while it is not given to us to choose what kind of universe we live
in now, we do have the freedom to improve our strategy for continued 
existence. It is of course hubris to believe that humans can at this point
foresee all the ideas that all future life
forms will come up with to save themselves.  In the future, there may
be many ways to work around the basic
premises we have here assumed. We list here a few of the
ones one can imagine.    Perhaps someday one can find a way
to create and use wormholes. Then we could either bring in energy
from far distant points in the universe for our use,
or we could travel to some other more congenial place in the 
universe where there are still sufficient resources for 
our consumption.  Another alternative
would be to create a universe in a lab, along the lines
of suggestions made by Guth and Farhi \cite{guthfarhi},
and then move into it. Future beings are likely to apply new
technology and sophistication, far beyond anything
we can anticipate, towards these questions of survival.

\section*{Acknowledgments}

KF acknowledges support from the Department of Energy via the University
of Michigan as well as the Institute for Particles, Strings,
and Astroparticle Physics at Columbia University. WHK is supported by ISCAP 
and  the Columbia University Academic Quality Fund. ISCAP gratefully 
acknowledges the generous support of the Ohrstrom Foundation. We thank
Glenn Starkman and Edward Baltz for helpful conversations.

\end{document}